\documentclass[3p,times]{elsarticle}

\usepackage{ecrc}

\volume{00}

\firstpage{1}

\journalname{Nuclear Physics A}

\runauth{}

\jid{procs}

\usepackage{amssymb}

\usepackage[figuresright]{rotating}

\begin{document}

\begin{frontmatter}



\dochead{}

\title{Non-photonic electron-hadron correlations and
non-photonic electron $v_2$ at STAR/RHIC}


\author{Gang Wang (for the STAR Collaboration)}

\address{Department of Physics and Astronomy, Unversity of California, Los Angeles, U.S.A.}

\begin{abstract}
Non-photonic electrons ($p_T >$ 3 GeV/$c$) represent the directions of the parent heavy quarks, 
and their angular correlations with charged hadrons provide a good tool to study the relative contributions 
from $D$ and $B$ meson decays in p+p collisions, as well as the 
flavor-dependence of the energy loss mechanism in A+A collisions.
We review the disentanglement of charm and bottom contributions 
to non-photonic electrons in p+p collisions at $\sqrt{s} = 200$ GeV at RHIC. 
$B$ decay contribution is approximately $50\%$ at the electron transverse momentum 
of $p_{T} > 5$ GeV/$c$ from STAR results. Incorporating the energy loss ($R_{AA}$) information of 
non-photonic electrons, the result indicates a $B$ meson suppression at high $p_T$ in heavy ion collisions. 
We also present non-photonic electron-hadron correlations in d+Au collisions as a baseline reference for the investigation
of the heavy quark energy loss in A+A collisions.
The interactions beteen heavy quarks and the QCD medium can lead to a non-zero 
 ``elliptic flow" parameter ($v_2$) of $D$ and $B$ mesons 
which can be studied through $v_2$ of non-photonic electrons.
Preliminary $v_2$ measurements for non-photonic electrons are shown to be lower than $v_2$ of charged hadrons 
(or $K_S^0$, $\Lambda$) for 200 GeV Au+Au collisions.
\end{abstract}


\end{frontmatter}


\section{Introduction}
\label{}
The ``non-photonic" electrons ($e_{non\gamma}$) from semi-leptonic decays of $D$ and $B$ mesons for $p_T$ up
to 9 GeV/$c$ in central Au+Au collisions have been observed with yields suppressed to a similar level
to that of light quarks at the Relativistic Heavy Ion Collider (RHIC) \cite{bib1,bib2}.
The suppresion was unexpected due to the ``dead cone effect"\cite{bib3}.
Fixed-Order-Next-to-Leading-Log (FONLL) pQCD calculations predict that 
the $B$-decay contribution is significant at and above $p_T$ of $4 - 5$ GeV/$c$,
though theoretical uncertainties are large\cite{bib4}. 
Measuring the bottom quark contribution to non-photonic electron yields in p+p
collisions is important in order to understand the production of heavy quarks and
further help to investigate the energy loss of heavy quarks in A+A collisions.

When high-$p_T$ partons lose a significant amount of energy traversing the dense QCD medium created in central Cu+Cu or
Au+Au collisions, their azimuthal correlations with low-$p_T$ hadrons are modified,
showing a broad or even double-peak structure on the away-side di-hadron correlation~\cite{bib5,bib6,bib7}.
Non-photonic electrons represent well the directions of the parent $D$ or $B$ mesons
when electron $p_T >$ 3 GeV/$c$, and in the opposite direction we expect another
heavy quark traversing the medium leaving an imprint on the away-side $e_{non\gamma}-h$ correlation.
A similar broadness has been observed on the away-side $e_{non\gamma}-h$  correlations in
200 GeV Au+Au and Cu+Cu collisions~\cite{bib8}, and the counterpart study in d+Au collisions
will provide a baseline for the energy loss
of heavy quarks in the hot and dense medium produced in central A+A collisions.

Heavy quarks are believed to be produced mostly via initial gluon fusion~\cite{bib9}
in relativistic heavy ion collisions, and propagation of
heavy quarks through the medium is a good probe to study the medium properties.
A non-zero $v_2$ of $D$ and $B$ mesons could arise from the interaction between the parent
heavy quarks and the medium, and could be approximated by non-photonic electron $v_2$.
Thus the $v_2$ measurement of non-photonic electrons will shed light on the extent of the thermalization of the collision system.

\section{Analysis and results}

\begin{figure}
\begin{minipage}[c]{0.48\textwidth}
\center
\includegraphics[width=\textwidth]{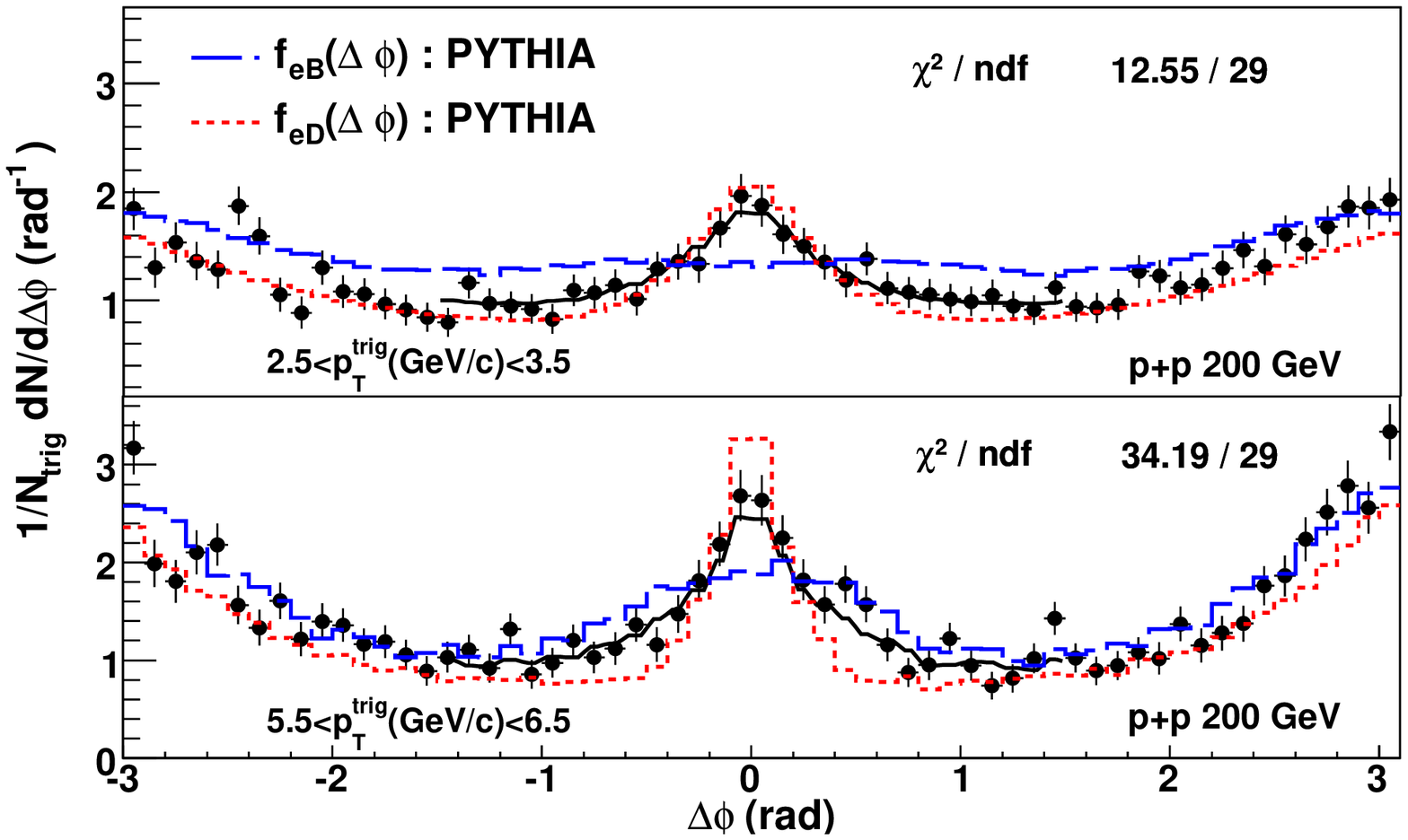}
  \caption{Distributions of the azimuthal angle between non-photonic electrons and charged hadrons normalized
per non-photonic electron trigger. The trigger electron has (top) $2.5 < p_T < 3.5$ GeV/$c$ and 
(bottom) $5.5 < p_T < 6.5$ GeV/$c$~\cite{bib10}. The curves represent PYTHIA calculations for $D$
(dotted curve) and $B$ (dashed curve) decays~\cite{bib11}. The fit result is shown as the black solid curve.}
\label{fig:eh_pp}
\end{minipage}
\begin{minipage}[c]{0.48\textwidth}
\includegraphics[width=\textwidth]{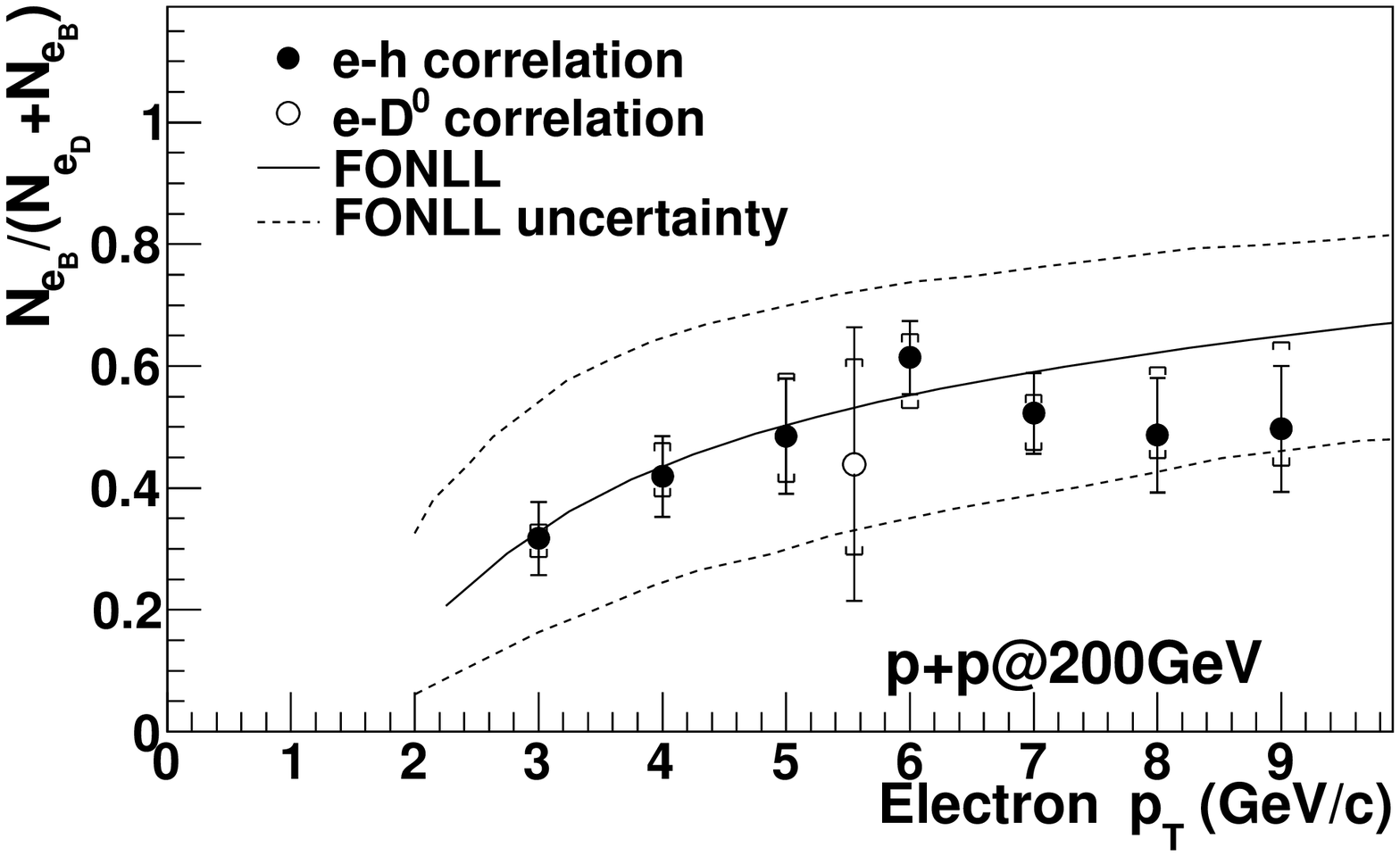}
  \caption{Transverse momentum dependence of the relative contribution from $B$ mesons to the non-photonic electron
yields~\cite{bib10}. 
Error bars are statistical and brackets are systematic uncertainties. The solid curve is the FONLL calculation~\cite{bib4}. 
Theoretical uncertainties are indicated by the dashed curves.}
\label{fig:B_contribution}
\end{minipage}
\end{figure}

In the analysis related to non-photonic electrons, the hadron contamination and the photonic background were removed.
The photonic background mainly comes from photon conversions in the detector material and Dalitz decays, 
dominantly from $\pi^0$.
The photonic electrons were identified by pairing electrons with oppositely charged partner tracks, 
determining the conversion or decay vertex, and calculating the invariant mass of $M_{e^+e^-}$~\cite{bib12}.
Figure~\ref{fig:eh_pp} gives the example of the measured $e_{non\gamma}-h$ correlations together with PYTHIA
calculations showing the different near-side widths when triggering on electrons from $D$ or $B$ mesons~\cite{bib11}.
The width difference enables us to carry out a combined fit of PYTHIA $D/B$ calculations
to the measured correlations with the only free parameter to be the relative $B$ contribution, $r_B$.
The fit results are shown in Fig.~\ref{fig:B_contribution} as a function of $e_{non\gamma}$ $p_T$.
$r_B$ increases with $p_T$ and reaches approximately 0.5  around $p_T = 5$ GeV/$c$,
consistent with the FONLL calculation~\cite{bib4} within theoretical and experimental uncertainties.
We also estimated that the inclusion of $e_{non\gamma}-h$ correlation from $J/\psi$ decays
will not alter $r_B$ significantly within statistical and systematic errors.

\begin{figure}
\center
\vspace{-0.2cm}
\includegraphics[width=0.55\textwidth]{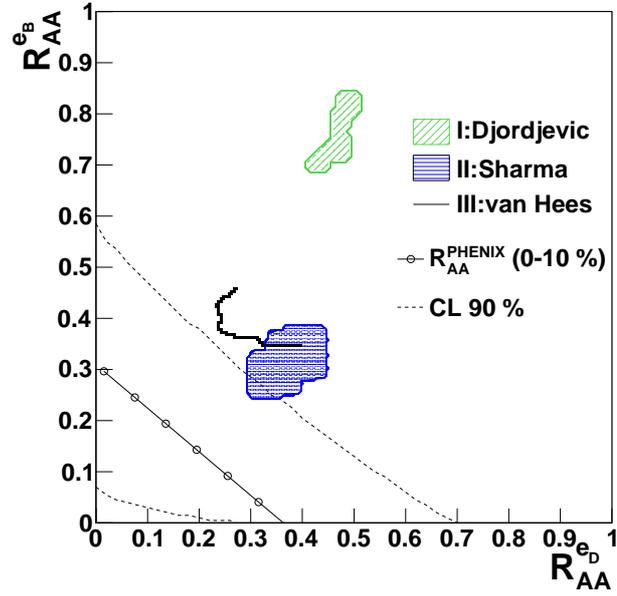}
\vspace{-0.3cm}
\caption{Confidence level contours for nuclear modification factor $R_{AA}$ for electrons 
from $D$ ($R^{e_D}_{AA}$) and $B$ ($R^{e_B}_{AA}$) meson decays and determined by combining 
the $R_{AA}$ results and the $r_B$ measurement for $p_T > 5$ GeV/$c$~\cite{bib10}. 
Three different models of $R_{AA}$ for $D$ and $B$ are described in the text.}
\label{fig:RAAeB}
\end{figure}

Neglecting contributions from charmed baryons, $e_{non\gamma}$ $R_{AA}$~\cite{RAA} is given by
\begin{equation}
R_{AA}^{non\gamma} = (1-r_B)R_{AA}^{e_D} + r_B R_{AA}^{e_B}, 
\label{eqn1}
\end{equation}
where $R_{AA}^{e_D}$ ($R_{AA}^{e_B}$) is the $R_{AA}$ for electrons from $D$ ($B$) mesons~\cite{bib10}.
With the current measurement of $r_B$ and the $R_{AA}^{non\gamma}$ measurement from PHENIX~\cite{bib13},
we picture Eq.~\ref{eqn1} in Fig.~\ref{fig:RAAeB} for $p_T > 5$ GeV/$c$, where the solid line represents the 
most probable values for $R_{AA}^{e_D}$ and $R_{AA}^{e_B}$, and the dashed lines, the $90\%$
Confidence Limit. This result indicates that $B$ meson yields are suppressed
at high $p_T$ in heavy ion collisions.
From comparison with model calculations (Fig.~\ref{fig:RAAeB}), we find that
those considering suppression of the heavy flavor hadrons due to
dissociation (model II~\cite{bib15}) or elastic scattering (model III~\cite{bib16})
processes are consistent with the measurements,
while the model considering only b-quark energy loss in the dense medium
(model I~\cite{bib14}) is inconsistent with the data.

\begin{figure}
\center
\vspace{-0.5cm}
\includegraphics[width=0.85\textwidth]{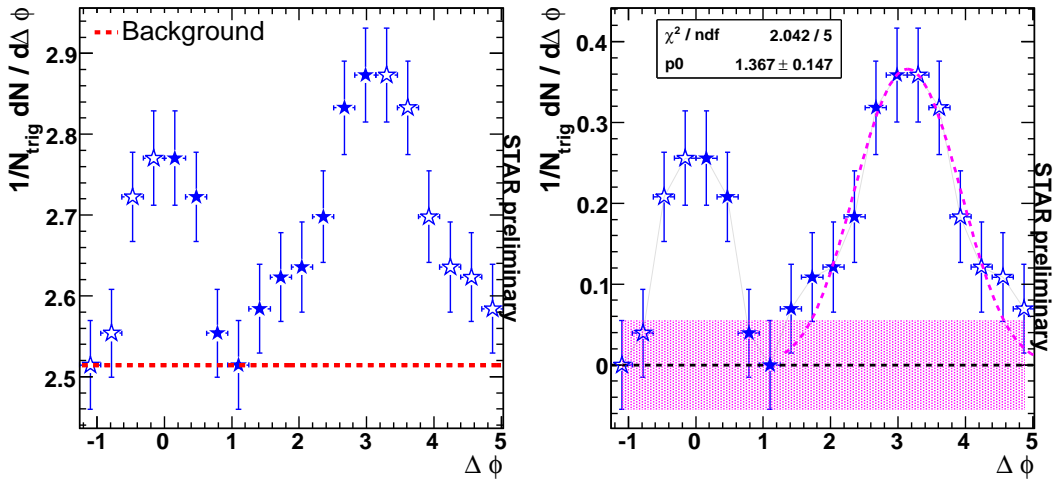}
\vspace{-0.3cm}
\caption{The azimuthal angular correlation between non-photonic electrons ($3<p_T<6$ GeV/$c$)
and hadrons ($0.15<p_T<0.5$ GeV/$c$) for 200 GeV d+Au collisions. 
The open points are reflections.
The left (right) panel shows the correlation before (after) background subtraction, 
with a dashed fitting curve from PYTHIA expectations on the away side. 
The error bars are statistical, and the error band around zero shows the systematical uncertainty from ZYAM~\cite{bib18}.}
\label{fig:eh_dAu}
\end{figure}

To study the jet-medium interactions in A+A collisions, d+Au collisions provide a better baseline
reference than p+p collisions, since ``cold nuclear matter" effects~\cite{bib17} such as nuclear
absorption and shadowing are accounted for in d+Au collisions. Fig.~\ref{fig:eh_dAu} shows $e_{non\gamma}-h$
correlations for RHIC run2008 200 GeV d+Au collisions, with the same $p_T$ ranges for the trigger and associated particles
as in Ref\cite{bib8}. 
To enhance the statistics, the measured correlations are folded into $[0, \pi]$,
and the data points beyond are reflections.
On the away side there is a single peak, and the correlation structure after background subtraction 
can be well described by PYTHIA calculations for p+p collisions, indicating that the impact of cold nuclear matter effects
is not prominent on this analysis for these specific $p_T$ ranges.

\begin{figure}
\center
\includegraphics[width=0.85\textwidth]{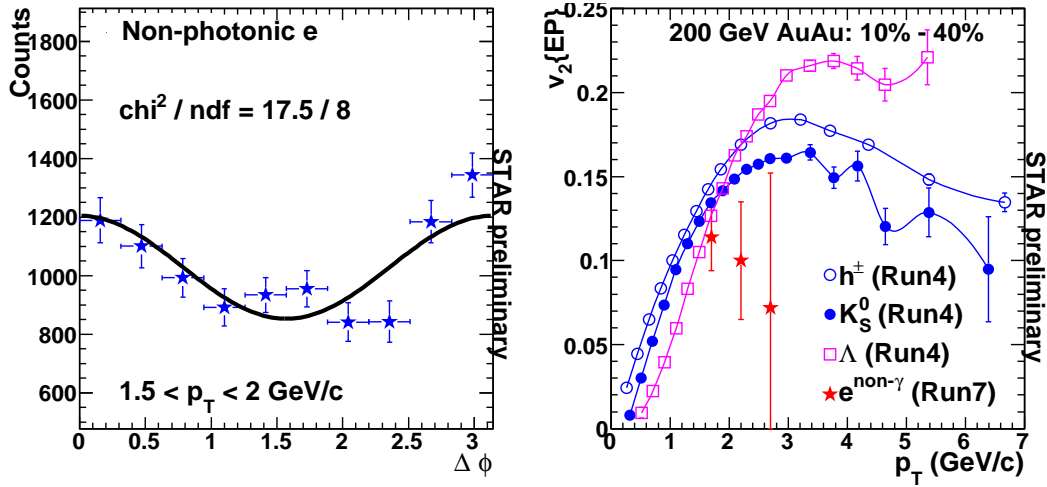}
\vspace{-0.3cm}
\caption{The left panel demonstrates the measurement of non-photonic electron $v_2\{EP\}$,
and the right panel shows the $v_2$ comparison between charged hadrons, $K_S^0$, $\Lambda$~\cite{bib20} and non-photonic electrons. 	}
\label{fig:v2}
\end{figure}

The left panel of Fig.~\ref{fig:v2} illustrates the extraction of $e_{non\gamma}$ $v_2$ via its angular azimuthal
correlation with the event plane~\cite{bib19} for $10-40\%$ RHIC run2007 200 GeV Au+Au collisions.
The retrieved $v_2$ parameter from the fit was corrected with the event plane resolution
and for the granularity effect due to the finite $\Delta \phi$ bin width, and the fit error was multiplied by the square root of
$\chi^2/ndf$ to compensate for the fitting quality.
The right panel shows the comparison of $v_2$ between
$e_{non\gamma}$ and charged hadrons ($K_S^0$, $\Lambda$)~\cite{bib20}. With the large statistical errors, $e_{non\gamma}$
$v_2$ is found to be systematically lower than that of hadrons, $K_S^0$ or $\Lambda$.
The feasibility of the heavy flavor tagged correlation analysis has been demonstrated,
and we expect a more complete study of this type based on the RHIC run2010 data
with much higher statistics and lower detector material budget.






\end{document}